\documentclass[sigconf,preprint]{acmart}

\usepackage{algorithmic}  
\usepackage{algorithm} 
\usepackage[algo2e]{algorithm2e} 
\usepackage{pifont,dsfont}
\usepackage{makecell}
\usepackage{amsfonts}
\usepackage{wrapfig,lipsum,booktabs}
\usepackage[export]{adjustbox}
\usepackage{xcolor,colortbl}
\usepackage{multirow}

\usepackage{graphicx}
\usepackage{subcaption}

\usepackage{refcount}
\usepackage{amsfonts}
\usepackage{caption}

\newcommand{\hide}[1]{}
\usepackage{xcolor,colortbl}

\newtheorem*{pro-stat}{Problem Definition}

\newcommand{\myModel}{{\textsc{MixDemo}}}

\AtBeginDocument{%
  \providecommand\BibTeX{{%
    \normalfont B\kern-0.5em{\scshape i\kern-0.25em b}\kern-0.8em\TeX}}}

\copyrightyear{2023} 
\acmYear{2023} 
\setcopyright{acmlicensed}\acmConference[WWW '23]{Proceedings of the ACM Web Conference 2023}{May 1--5, 2023}{Austin, TX, USA}
\acmBooktitle{Proceedings of the ACM Web Conference 2023 (WWW '23), May 1--5, 2023, Austin, TX, USA}
\acmPrice{15.00}
\acmDOI{10.1145/3543507.3583316}
\acmISBN{978-1-4503-9416-1/23/04}

\begin{document}

\title{Mixture of Demonstrations for Textual Graph Understanding and Question Answering}

\author{Yukun Wu, Lihui Liu}
\email{yukun.wu.mail@gmail.com, hw6926@wayne.edu}
\affiliation{%
  \institution{Independent Researcher, Wayne State University}
  \city{Detroit}
  \country{USA}
}






\begin{abstract}

Textual graph-based retrieval-augmented generation (GraphRAG) has emerged as a powerful paradigm for enhancing large language models (LLMs) in domain-specific question answering. While existing approaches primarily focus on zero-shot GraphRAG, selecting high-quality demonstrations is crucial for improving reasoning and answer accuracy. Furthermore, recent studies have shown that retrieved subgraphs often contain irrelevant information, which can degrade reasoning performance.
In this paper, we propose \myModel, a novel GraphRAG framework enhanced with a Mixture-of-Experts (MoE) mechanism for selecting the most informative demonstrations under diverse question contexts. To further reduce noise in the retrieved subgraphs, we introduce a query-specific graph encoder that selectively attends to information most relevant to the query.
Extensive experiments across multiple textual graph benchmarks show that \myModel\ significantly outperforms existing methods.

\end{abstract}




\maketitle

\section{Introduction}

Large language models (LLMs) have achieved remarkable success in recent years. Yet, most LLMs are trained on open-domain data before fixed cut-off dates ~\cite{gpt2}, which inevitably limits their performance when facing domain-specific questions due to outdated or missing knowledge. To address this issue, Retrieval-Augmented Generation (RAG) ~\cite{replug, graphrag} has emerged as a promising solution, where relevant knowledge is retrieved and incorporated to help LLMs generate accurate responses. 
Most existing GraphRAG methods ~\cite{graphrag,gretriever} rely on a simple yet strong assumption, that the generated response should include all relevant facts retrieved from the graph. However, this assumption overlooks a critical problem: the retrieved textual graph might contain unnecessary noise or irrelevant information.
Consider Figure ~\ref{example}(a), where a user asks “What is a good source of nutrients for a mushroom?” While the textual graph contains correct 'a cut peony', it also contains useless information 'a flying eagle' which may mislead the model. 
How to mitigate low-quality information from the retrieved textual graph to effectively guide answer generation is a problem.
Furthermore, existing GraphRAG methods mainly utilize zero shot learning. Selecting
high-quality demonstrations is crucial for improving reasoning and answer accuracy.

In this paper, we propose a GraphRAG framework that focuses on enhancing both textual graph understanding and QA performance by learning to select high-quality demonstrations and query specific information learning. 
Instead of assuming all retrieved subgraph content is useful, our approach adaptively identifies the most informative and context-relevant node-level and edge-level evidence. 
Specifically, we design a specialized, query-specific GraphEncoder to model the complex interactions between nodes within the retrieved subgraphs. 
This encoder generates a dense graph prompt embedding that captures relational patterns and serves as a bridge between structured knowledge and the LLM’s input space. 
Additionally, we incorporate a Mixture-of-Experts (MoE) ~\cite{deepseekmoe} paradigm to select the most informative demonstrations to enhance in-context learning. 
We evaluate our method on the GraphQA benchmark. Extensive experiments show that our model significantly outperforms existing baselines.

\section{Problem Definition}\label{problem-definition}

We study the task of answering queries over a \emph{textual graph} using large language models (LLMs). A textual graph \( G = (V, E) \) contains natural-language content on both nodes and edges. Given a query \( q \), the goal is to generate an answer \( a_{\mathrm{gen}} \) by retrieving a relevant subgraph \( S \subseteq G \) and using it to guide LLM-based reasoning.

We adopt \emph{in-context learning} (ICL), where a set of demonstration examples \(\mathcal{D} = \{(q_i, a_i)\}\) is prepended to the query to prompt the LLM. At test time, a subset \(\mathrm{Sub}(D)\) is selected from \(\mathcal{D}\) to construct the prompt. The LLM then predicts:
\[
\hat{a} = \arg\max_a P_{\mathrm{LM}}(a \mid \mathrm{Sub}(D), q).
\]

\paragraph{Final Problem Definition.}
Our goal is to answer a query \(q\) on a textual graph \(G\) by: (1) retrieving a relevant subgraph \(S\), and (2) prompting an LLM via ICL with demonstrations \((S_i, q_i, a_i)\) to generate an accurate answer.

\section{Proposed Method}\label{overview}

\begin{figure*}[h!]
    \centering
    \includegraphics[width=1.0\textwidth]{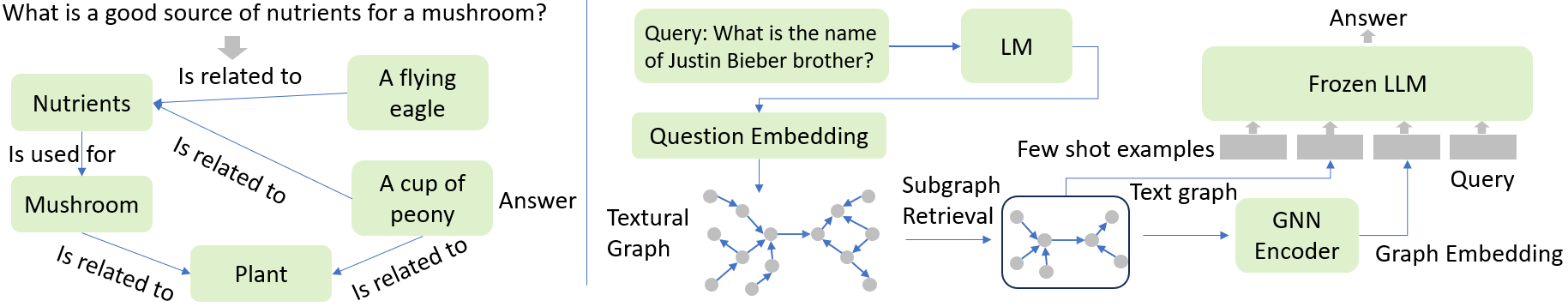} 
    \vspace{-1\baselineskip}
    \caption{(a) An example of a retrieved subgraph. (b) Overview of \myModel.}
    \label{example} 
\end{figure*}

Our method tackles two key challenges in textual graph question answering: noisy subgraphs and limited examples. Given a query $q$ and textual graph $G = (V, E)$, we retrieve a subgraph $S$ using GRetriever~\cite{gretriever}. To reduce noise in $S$, we apply a query-aware graph attention network that emphasizes relevant nodes and edges. To further enhance reasoning, we use few-shot in-context learning with selected $(q_i, S_i, a_i)$ examples, enabling the model to learn from both textual and structural patterns. The framework is shown in Figure ~\ref{example}.

\subsection{Subgraph Retrieval}

Given a query $q$, we use a pretrained language model (Sentence-BERT ~\cite{sentencebert}) $\mathrm{LM}(\cdot)$ to encode $q$, nodes, and edges in the textual graph $G = (V, E)$:

{\small
\begin{align}
\mathbf{z}_q &= \mathrm{LM}(q), 
\mathbf{z}_{v_i} = \mathrm{LM}(v_i), 
\mathbf{z}_{e_{i,j}} = \mathrm{LM}(e_{i,j})
\end{align}
}%

We compute cosine similarities between $\mathbf{z}_q$ and all node/edge embeddings and retrieve the top-$k$ nodes $V_k$ and edges $E_k$ with highest similarity. A connected subgraph is then constructed using the Prize-Collecting Steiner Tree (PCST) algorithm~\cite{gretriever}. Each retrieved item is assigned a prize based on its rank, and PCST selects a subgraph $S$ that maximizes total prize while minimizing edge cost:

{\small
\begin{align*}
S = \underset{S \subseteq G}{\operatorname{argmax}} ( \sum_{v_i \in V_k} \text{prize}(v_i) + \sum_{e_{i,j} \in E_k} \text{prize}(e_{i,j}) - c \cdot |E_S| ).
\end{align*}
}%

\subsection{Demonstration Retrieval}

Building on our subgraph retrieval method, we introduce an approach to select informative few-shot examples for improved language model reasoning.
Given a query $q$, we convert its retrieved subgraph $S$ into text using $\text{textualize}(\cdot)$, which flattens all node and edge attributes. The textualized graph is concatenated with $q$ to form the prompt $x$: $x = \mathrm{textualize}(S)||q$.

While nearest-neighbor retrieval in embedding space is common, it may miss globally relevant examples. Inspired by recent work~\cite{onepromptisnotenough}, we instead cluster demonstrations by semantic similarity and select a representative from each cluster for more diverse and complementary examples.
Specifically, We apply K-means clustering to partition the example pool
$\mathcal{D} = {(S_i, q_i, a_i)}_{i=1}^n$
into $C$ clusters ${C_1, C_2, \dots, C_C}$, treating each cluster as an expert. Clustering is performed on the Sentence-BERT embeddings of the augmented prompts $x_i = \mathrm{textualize}(S_i) || q_i$, following prior work~\cite{gretriever}.

To adaptively determine the optimal number of clusters $C$, we minimize a regularized objective that balances within-cluster variance and model complexity:
\begin{equation}
C^* = \min_C \sum_{k=1}^{C} \sum_{x_i \in C_k} |f(x_i) - \boldsymbol{\mu}_k|_2^2 + \lambda C,
\end{equation}
where $f(\cdot)$ denotes the embedding function, $\boldsymbol{\mu}_k$ is the centroid of cluster $C_k$, and $\lambda$ controls the regularization strength.

At inference time, a test query $q$ is augmented into $x_q = \mathrm{textualize}(S_q) || q$, embedded via $f(x_q)$, and assigned to its closest expert based on cosine similarity with cluster centroids:
\begin{equation}
c(q) = \arg\max_{i = 1, \dots, C^*} \cos(f(x_q), \boldsymbol{\mu}_i).
\end{equation}
The selected expert then provides a set of representative demonstrations, which are combined with the input to generate the model’s final prediction.

\subsection{Noise Mitigation}

Building on our retrieval and demonstration selection framework (Section 3.2), we now describe how the final answer is generated using the retrieved subgraphs. Specifically, we encode each subgraph $S$ with a \textit{GraphEncoder} that transforms its structural and semantic information into a graph-prompt representation. This encoding serves two purposes: (1) preserving relational patterns critical to the query, and (2) filtering irrelevant information through query-sensitive attention, thereby constructing an optimized input for the language model.

Prior methods like G-Retriever use GCNs ~\cite{gcn} or GATs~\cite{gat}, but these suffer from over-smoothing~\cite{oversmooth}, making node embeddings indistinguishable, an issue in our setting where retrieved subgraphs mix relevant and noisy content. Effective encoding thus requires \textit{query-aware} representations that selectively highlight important nodes. For instance, given a query, the encoder should emphasize \texttt{a cut peony} and ignore irrelevant nodes like \texttt{a flying eagle}, even if structurally nearby.
To achieve this, we design a query-conditioned GNN where both message passing and node interactions are modulated by the input query $q$. Specifically, we redefine the edge weight $\zeta_{e_{i,j}}^{(l)}$ using a query-aware attention mechanism, allowing the model to focus on the most informative edges.
At each layer $l$, the attention weight $\zeta_{e_{i,j}}^{(l)}$ is computed by:
\begin{align}
\alpha_{v_i}^{(l)} &= \texttt{LINEAR}\left(\texttt{CONCAT}(z_{v_i}^{(l)}, q)\right) \\
\beta_{v_j}^{(l)} &= \texttt{LINEAR}\left(\texttt{CONCAT}(z_{v_j}^{(l)}, q)\right) \\
\gamma_{e_{i,j}} &= \texttt{LINEAR}\left(\texttt{CONCAT}(z_{e_{i,j}}, q)\right) \\
\zeta_{e_{i,j}}^{(l)} &= \texttt{tanh}\left(\alpha_{v_i}^{(l)} + \gamma_{e_{i,j}} - \beta_{v_j}^{(l)}\right)
\end{align}
where $\alpha_{v_i}^{(l)}$, $\beta_{v_j}^{(l)}$, and $\gamma_{e_{i,j}}$ are learned {query-conditioned} node/edge embeddings. $\zeta_{e_{i,j}}^{(l)}$ serves as the attention weight for message passing 
along edge $e_{i,j}$. Similarly, messages are generated using query-conditioned features:
\begin{equation}
\text{msg}_{e_{i,j}}^{(l)} = \texttt{LINEAR}\left(\texttt{CONCAT}(z_{v_i}^{(l)}, z_{v_j}^{(l)}, z_{e_{i,j}}, q)\right)
\end{equation}
and aggregated via attention weights:
\begin{equation}
z_{v_j}^{(l+1)} = \frac{1}{d_{v_j}}\sum_{v_i \in \mathcal{N}(v_j)} \zeta_{e_{i,j}}^{(l)} \cdot \text{msg}_{e_{i,j}}^{(l)}
\end{equation}

Unlike standard GCNs that apply static, query-agnostic filters, our GNN conditions node interactions and message passing on the query \( q \). Specifically, we redefine edge weights \( \zeta_{e_{i,j}}^{(l)} \) using a query-aware attention mechanism, allowing the model to emphasize task-relevant nodes and edges.

After \( L \) layers, each node \( v_j \in S \) has an embedding \( z_{v_j}^{(L)} \), which we mean-pool to obtain the subgraph representation:
$z_S = \texttt{POOL}(z_{v_j}^{(L)})$.
This is projected into the LLM embedding space using an MLP:
$p_\text{graph} = \texttt{MLP}(z_S)$,
serving as the graph prompt.

To incorporate multiple demonstration subgraphs \( \{z_d\}_{d \in D_R} \), we 
use query-based relevance weighting:

{\small
\begin{align}
\lambda(q, z_d) = \frac{e^{s(q, z_d)}}{\sum_{d'} e^{s(q, z_{d'})}}, 
z_{\text{final}} = \sum_{d=0}^N \lambda(q, z_d) z_d,
\end{align}
}

where \( z_0 \equiv z_{\text{current}} \).

\textbf{Generating responses.} We prepend task-specific instructions and tokenize all inputs:
$q = \texttt{tokenize}(q)$, $p_\text{demo} = \texttt{tokenize}(p_\text{demo})$, $p_\text{text-graph} = \texttt{tokenize}(p_\text{text-graph})$,
then feed the combined sequence into the frozen LLM:
\begin{equation}
a_\text{gen} = \texttt{LLM}(\texttt{CONCAT}(p_\text{demo}, p_\text{graph}, p_\text{text-graph}, q)),
\end{equation}
yielding the final answer \( a_\text{gen} \).

\label{sec:method}

\section{Experiments}\label{experiments}

\begin{table}[ht]
\centering
\caption{Statistics of datasets. FB means FreeBase.}
\small
\label{tab:dataset_statistics}
\begin{tabular}{|c|c|c|c|c|}
\hline
\textbf{Dataset} & \textbf{ExplaGraphs} & \textbf{SceneGraphs} & \textbf{WebQSP}  \\ \hline
\#Graphs          & 2,766               & 100,000             & 4,737                    \\ \hline
Average \#Nodes   & 5.17                & 19.13               & 1370.89                  \\ \hline
Average \#Edges   & 4.25                & 68.44               & 4252.37                 \\ \hline
Node Attribute    &  concepts & Object attributes    & Entities in FB  \\ \hline
Edge Attribute    &  relations & Spatial relations   & Relations in FB  \\ \hline
Task              &  reasoning & Scene graph QA       & KGQA             \\ \hline
Evaluation metrics & Accuracy            & Accuracy            & Hit@1              \\ \hline
\end{tabular}
\end{table}

\begin{table*}[ht]
\centering
\caption{Performance comparison for different methods (\%).}
\vspace{-0.7\baselineskip}  
\small
\label{tab:performance_comparison_A_side}
\begin{tabular}{|c|c|c|c|}
\hline
\textbf{Dataset (Metrics)} & \textbf{ExplaGraphs (ACC)} & \textbf{SceneGraphs (ACC)} & \textbf{WebQSP (Hit@1)} \\ \hline
Zero-shot                   & 56.50                      & 39.74                      & 41.06                    \\ 
Zero-CoT (Kojima et al., 2022) & 57.04                  & 52.60                      & 51.30                    \\ 
CoT-BAG (Wang et al., 2024)  & 57.94                      & 56.80                      & 39.60                    \\ 
KAPING (Baek et al., 2023)   & 62.27                      & 43.75                      & 52.64                    \\ 
Graph-based Inference        & 33.93                      & 42.17                      & 47.22                    \\ 
Frozen LLM + Prompt Tuning (PT) & 58.98                  & 63.72                      & 54.11                    \\ 
GraphToken (Perozzi et al., 2024) & 85.08               & 49.03                      & 57.05                    \\ 
G-Retriever                  & 86.19                      & 80.86                      & 70.02                    \\ \hline
\myModel\                    & 87.31                      & 82.32                      & 71.36                    \\ \hline
\end{tabular}
\label{result_table}
\vspace{-1.3\baselineskip}  
\end{table*}

\noindent\textbf{Datasets.}  
We evaluate on the GraphQA benchmark~\cite{gretriever}, which includes ExplaGraphs, SceneGraphs, and WebQSP. Dataset stats are in Table~\ref{tab:dataset_statistics} in the Appendix.
\noindent\textbf{Metrics.}  
Following GRetriever, we use accuracy for ExplaGraphs and SceneGraphs, and Hit@1 for WebQSP, which allows multiple correct answers.
\noindent\textbf{Baselines.}  
We compare against inference-only methods (e.g., Zero-shot ~\cite{zeroshot}, CoT-BAG~\cite{CoT-BAG}, KAPING~\cite{KAPING}) and prompt-tuning methods (e.g., Prompt Tuning, GraphToken~\cite{GraphToken}, G-Retriever~\cite{gretriever}).




\subsection{Effectiveness of \myModel}

The results are summarized in Table~\ref{result_table}, which compares \myModel{} against all baseline methods. Overall, \myModel{} consistently achieves superior performance across all datasets. For example, it outperforms the strongest baseline, G-Retriever, by approximately 1.1\% on ExplaGraphs and 1.5\% on SceneGraphs. These improvements highlight the effectiveness of the proposed approach.
Additionally, we note that naively textizing the retrieved subgraph information and using it as direct input for LLMs often yields poor results, in most cases, performance is significantly degraded. This demonstrates the importance of properly encoding subgraph structural information and integrating it into LLMs.

\subsection{Ablation Study}

\begin{figure}[h]
\centering
\begin{subfigure}[t]{0.23\textwidth}  
  \includegraphics[width=\linewidth]{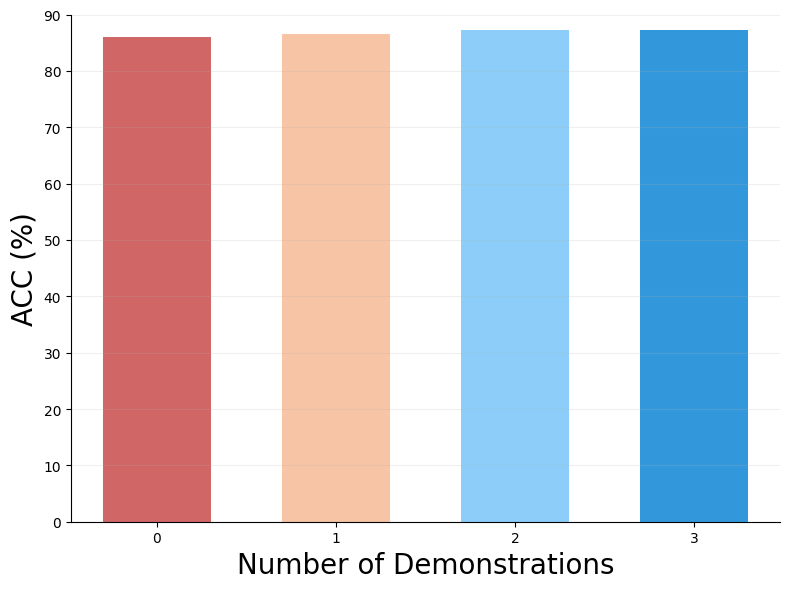}
  \caption{Study on Demo Number}
  \label{fig:pnnl1}
\end{subfigure}
\hfill  
\begin{subfigure}[t]{0.23\textwidth}  
  \includegraphics[width=\linewidth]{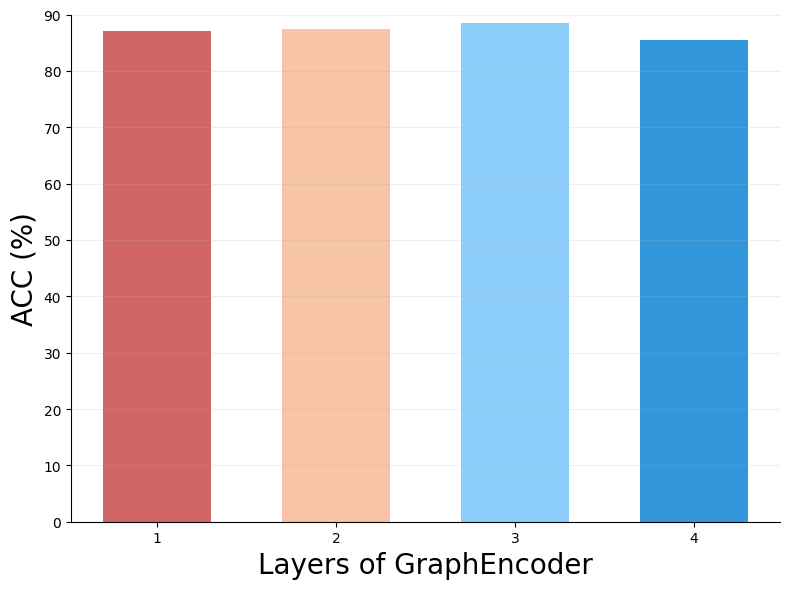}
  \caption{Study on GNN's Layers}
  \label{fig:pnnl2}
\end{subfigure}
\vspace{-0.5\baselineskip}  
\caption{Ablation study.}
\label{filtering}
\vspace{-1\baselineskip}  
\end{figure}



We first evaluate how the number of few-shot examples affects \myModel{}'s performance under zero-shot, 1-shot, 2-shot, and 3-shot settings. Since subgraphs in SceneGraphs and WebQSP exceed the LLM's input limit, we perform this study only on ExplaGraphs. As shown in Figure~\ref{fig:pnnl1}, 2-shot learning yields the best performance, with 2-shot and 3-shot results being nearly identical.

In the hyperparameter study, we assess how the number of GraphEncoder layers impacts performance. As shown in Figure~\ref{fig:pnnl2}, using three layers achieves the highest accuracy. Adding more layers offers no further improvement and can cause overfitting. These results highlight the importance of tuning encoder depth for optimal reasoning in \myModel{}.

\section{Related work}\label{related-work}

Recent work has highlighted Retrieval-Augmented Generation (RAG) ~\cite{gao-rag} as a powerful solution to mitigate key limitations of large language models (LLMs), particularly their tendency toward hallucinations in domain-specific or knowledge-intensive tasks. Current RAG methodologies can be broadly grouped into three paradigms. The simplest form, naive RAG ~\cite{ma-rag}, operates through a basic pipeline of indexing, retrieval, and generation. Building upon this foundation, advanced RAG systems incorporate optimizations during pre-retrieval, leveraging techniques like query transformation, expansion, and rewriting ~\cite{peng-rewriting,zheng-rewriting}, while post-retrieval enhancements often involve reranking strategies ~\cite{qin-rerank}. 

The Mixture of Experts (MoE) framework ~\cite{moe} has established itself as a fundamental paradigm in machine learning for developing adaptive systems and knowledge graph reasoning ~\cite{liu2019g,liu2021neural,liu2021kompare,liu2022joint,liu2022comparative,liu2022knowledge,liu2023knowledge,liu2023knowledge,hill2024ginkgo,liu2024logic,liu2024can,liu2024new,liu2024conversational,liu2025neural,liu2025few,wu2025improving,liu2025monte,liu2025hyperkgr,liu2025mixrag,liu2024knowledge,liu2026neural,liuneural,liu2026accurate,liu2026ambiguous,liu2026accurate,liu2026ambiguous,liu2025unifying,liu2026dynamic,liu2026accurate,liu2026symbolic,liu2025graph}. Initial work focused on traditional machine learning implementations ~\cite{NIPS1996_6c8dba7d}, with subsequent breakthroughs emerging through its integration with deep neural networks ~\cite{gcn}.
More recently, researchers have explored applying MoE approaches to in-context learning scenarios ~\cite{onepromptisnotenough}, demonstrating their potential to enhance large language model (LLM) performance. 

\section{Conclusion}\label{conclusion}

We present \myModel, a GraphRAG framework which leverages a Mixture-of-Experts demonstration selector and a query-aware graph encoder. By dynamically selecting contextually relevant demonstrations and filtering noisy subgraph information, our approach significantly improves answer accuracy and reasoning robustness across textual graph benchmarks. Experimental results validate that \myModel\ outperforms state-of-the-art baselines, demonstrating the importance of adaptive retrieval and noise reduction in GraphRAG systems.






\bibliographystyle{ACM-Reference-Format}
\bibliography{008reference,liu}



\end{document}